\documentclass[preprint,aps,superscriptaddress]{revtex4}
\usepackage{graphicx}
\usepackage{dcolumn}
\usepackage{bm}
\usepackage{times}

\begin{document}

\title{Nonlinear absorption and refraction in near-detuned rubidium vapor}

\author{C. F. McCormick, D. R. Solli, R. Y. Chiao}
\affiliation{Department of Physics, University of California, 
Berkeley, California 94720-7300}

\author{J. M. Hickmann}
\affiliation{Departamento de F\'{\i}sica, Universidade Federal de Alagoas, 
Cidade Universit\'{a}ria, 57072-970, Macei\'{o}, AL, Brazil}

\begin{abstract}
Using the z-scan technique, we have measured the self-induced absorptive and 
refractive nonlinear behavior of hot atomic rubidium vapor within the Doppler 
profile of the D2 line.  We observe large nonlinear amplitude and phase effects 
with only tens of microwatts of incident power.  Our results are in good 
agreement with numerical calculations based on an analytic model of a Doppler-
broadened two-level system.
\end{abstract}


\maketitle

Atomic vapors are interesting nonlinear optical materials because their 
nonlinear coefficients depend strongly on detuning, and nonlinear effects can be 
observed with very low cw laser power.  Nonlinear optics experiments in atomic 
vapors have included self-focusing, self-defocusing, self-trapping and self-
bending \cite{Grischkowsky1970, Grischkowsky1972, Bjorkholm1974, 
Swartzlander1988}; third-harmonic generation \cite{Young1971}; and four-wave 
mixing \cite{Harter1981}.  Several measurements of Kerr nonlinear coefficients 
have been reported in atomic vapors, for large detunings \cite{Lehmberg1976}, 
and some recent attention has focused on employing quantum coherence to control 
atomic nonlinear properties \cite{Wang2002}.

There has also been recent discussion concerning the role of the nonlinearity of 
atomic rubidium vapor in light-by-light guiding \cite{Truscott1999}.  
Theoretical work using three-level \cite{Kapoor2000} and five-level 
\cite{Andersen2001} models has been reported, both of which incorporate Doppler 
broadening numerically. The latter of these papers argues that the five-level 
model is necessary to accurately predict the self-action of a pump beam 
propagating in Rb vapor, and presents theoretical results on the index of 
refraction variation due to self-action.

In this work, we report a measurement of self-induced nonlinear absorption and 
refraction in hot atomic Rb vapor, within the Doppler profile of the D2 line 
($\lambda$ = 780 nm).  The nonlinearity is due to saturated atomic absorption, 
and only tens of microwatts are needed to get into the saturation regime for 
detunings within the Doppler profile.  Inhomogenous Doppler broadening 
complicates the form of the nonlinearity and a Kerr model is insufficient to 
describe it.  A fully analytic solution exists for the nonlinear behavior of a 
Doppler-broadened, two-level system, and it is reasonably tractable in the limit 
of small power-broadened linewidth compared with the Doppler width 
\cite{Close1967}.  Using this model we successfully predict the beam propagation 
and self-action effects in our experiment.  While a slight variant of this model 
has frequently been discussed in relation to nonlinear absorption in 
inhomogeneously broadened two-level systems, its application to self-action 
refractive effects has been much rarer \cite{textbooks}.

\begin{figure}
\centerline{\includegraphics{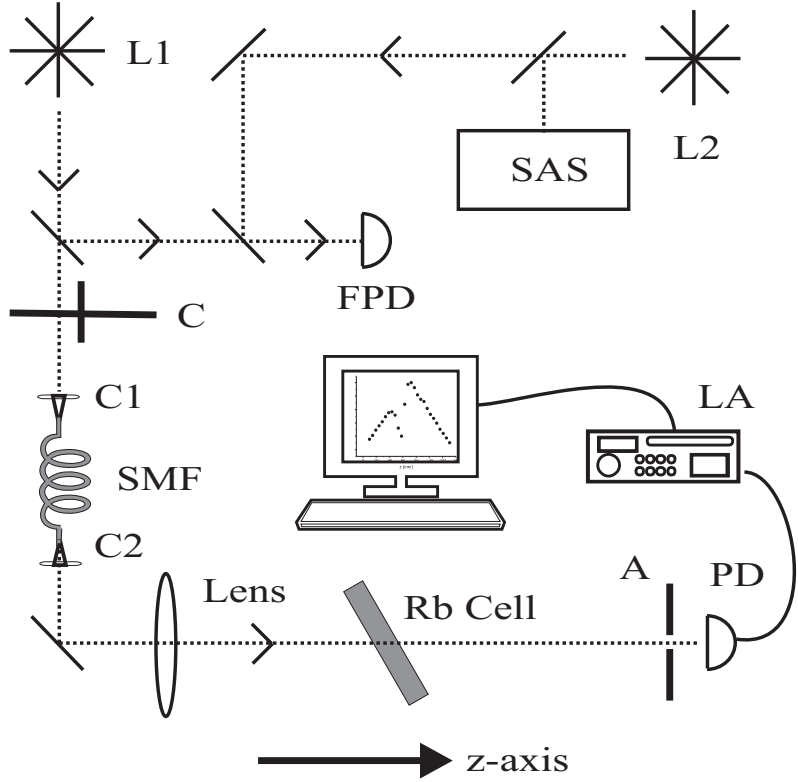}}
\caption{Experimental configuration for the z-scan measurement.  L1 and L2 = 
lasers 1 and 2; SAS = saturated absorption spectroscopy lock; C = chopper wheel; 
C1 and C2 = fiber coupling lenses; SMF = single mode fiber; A = aperture; PD = 
photodiode; LA = lock-in amplifier; FPD = fast photodiode.}
\label{configuration}
\end{figure}

Our experimental setup is a standard z-scan configuration \cite{Sheik-Bahae1989} 
(see Fig. \ref{configuration}).  The primary beam is provided by a 10 mW tunable 
cw diode laser with a bandwidth $\delta \nu \approx$ 300 kHz.  We measure its 
detuning to within 10 MHz by interfering it on a fast photodiode with a second 
diode laser locked using saturation absorption spectroscopy to the crossover 
resonance $5S_{1/2} \ F=3 \rightarrow 5P_{3/2} \ F'=3,4$ of $^{85}$Rb (in the D2 
line), and computing a real-time fast Fourier transform of the photodiode 
signal.  The multimode transverse profile of the experimental beam is 
regularized to a cylindrically symmetric, nearly-Gaussian shape by coupling it 
through a single-mode fiber with about 30\% efficiency.  The beam is focused by 
a lens to a Rayleigh range $z_{R} \approx$ 8 mm and is detected in the far-field 
using a photodiode. An aperture is placed in front of the photodiode, centered 
on the experimental beam. The intensity noise of the beam at the detector is 
less than 0.5\%.

We place a thin (L = 1 mm $\ll z_{R}$) optical vapor cell filled with 
natural-abundance Rb in the path of of the beam after the lens.  It has no 
magnetic shielding and experiences only the geomagnetic field.  The cell is 
heated to 78$^{\circ}$ C and tilted at 30$^{\circ}$ to prevent etalon effects.  
It is mounted on a 250 mm translation stage moved along the length of the 
experimental beam by a computer-controlled stepper motor.  To measure the noise 
in our system, we detune far off-resonance and record the transmission through 
the vapor cell as it is scanned 250 mm along the focused beam. For both fully 
open and 20\% fluence apertures these scans show less than $\pm$ 1\% 
transmission variation.  

\begin{figure}
\centerline{\includegraphics{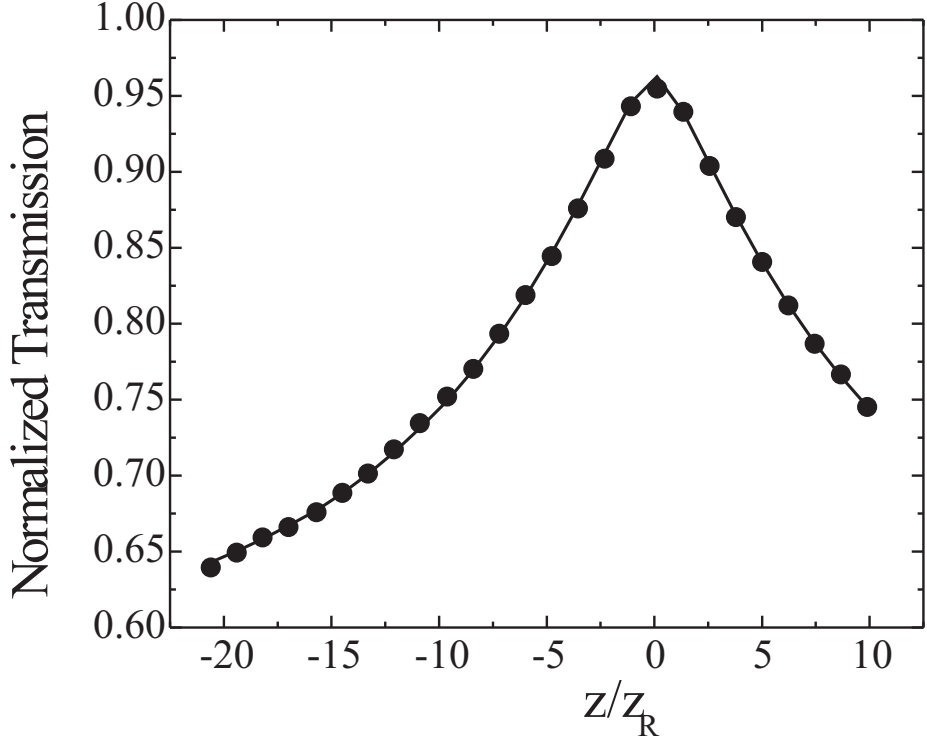}}
\caption{Open-aperture scan at 300 MHz to the blue of the $^{85}$Rb F=3 
$\rightarrow$ F' line, with fit from the Doppler-broadened two-level model.  The 
scan at 300MHz to the red is very similar.}
\label{openscan}
\end{figure}

\begin{figure}
\centerline{\includegraphics{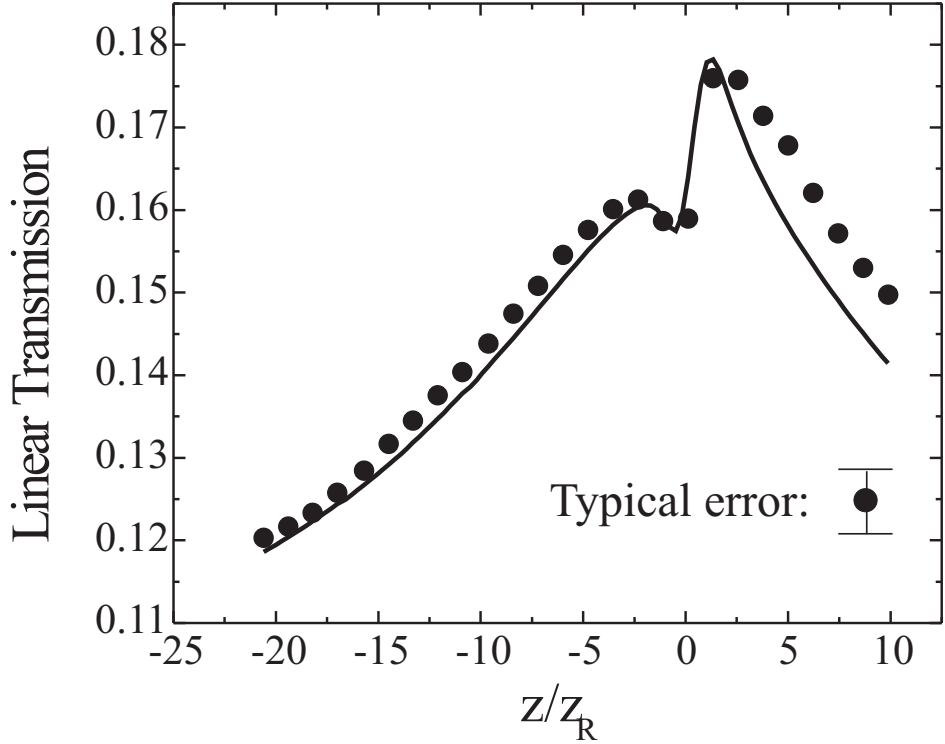}}
\caption{Closed-aperture scan at 300 MHz to the blue of the $^{85}$Rb F=3 
$\rightarrow$ F' line. The theory curve is calculated directly from the fit 
parameters found in the corresponding open scan, with no adjustable parameters. 
The typical systematic error mostly results from residual beam astigmatism and 
uncertainties about the aperture size.}
\label{closedscanblue}
\end{figure}

\begin{figure}
\centerline{\includegraphics{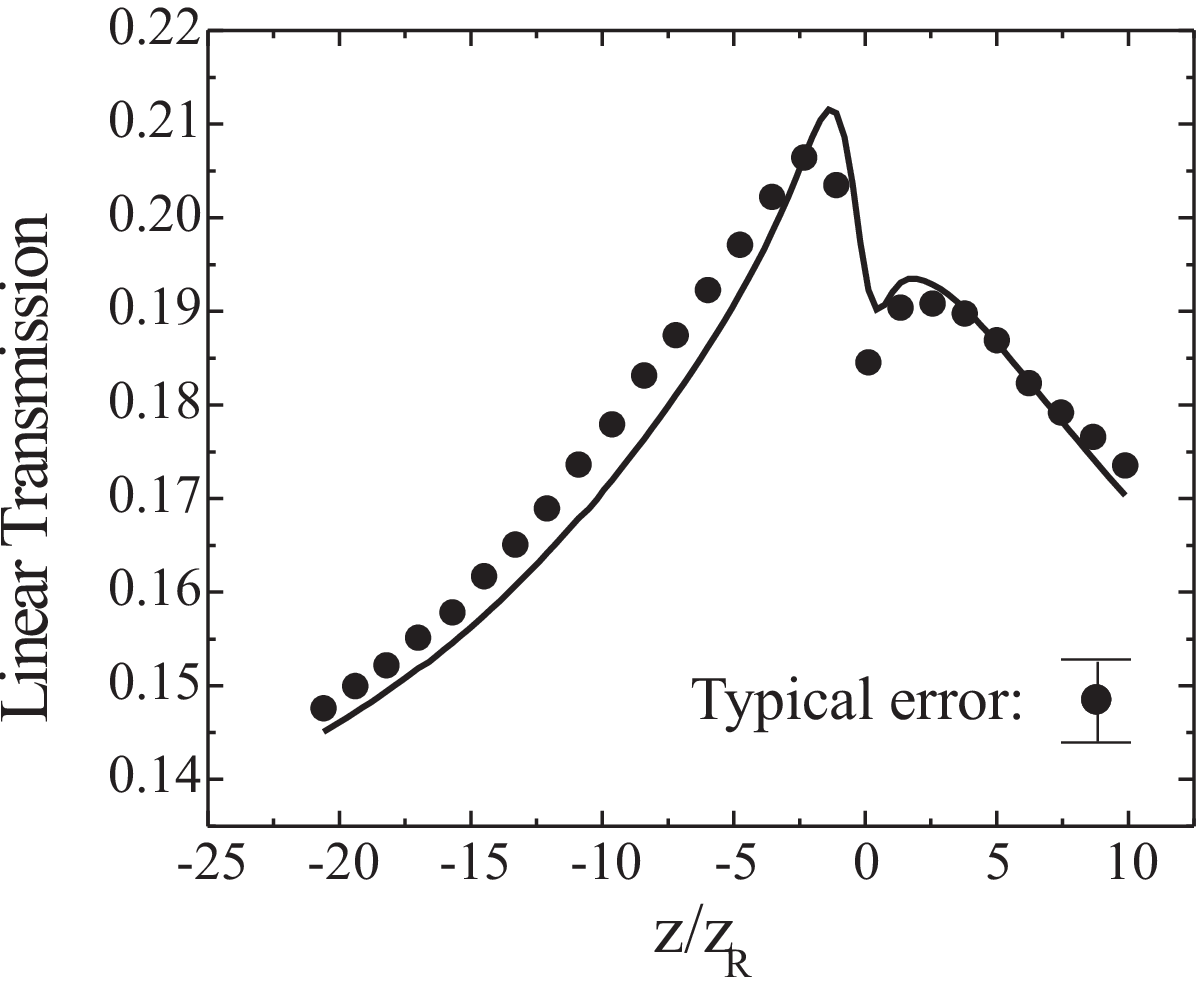}}
\caption{As above, for a detuning of 300 MHz to the red.}
\label{closedscanred}
\end{figure}

Using horizontal linear polarization and 45 $\mu$W of power, we performed scans 
with the aperture both fully open and closed to 20\% linear fluence, at 
detunings of $\pm$ 300 MHz from the $^{85}$Rb $F=3 \rightarrow F'$ transition 
(see Figs. \ref{openscan}, \ref{closedscanblue}, \ref{closedscanred}).  This 
detuning was chosen as representative because it is of the same order as the 
Doppler width of the transition, $\approx$ 380MHz. With the aperture open, there 
should be no refractive effects, and these scans display the qualitative 
behavior expected from a negative absorptive nonlinearity (i.e. absorption 
saturation).  With the aperture closed, the nonlinear phase imprinted by the 
atoms on the beam will change the far-field beam width and be detected as an 
asymmetric transmission profile with z, as we observe.  The symmetry of these 
scans is that of a self-focusing (-defocusing) nonlinearity for blue (red) 
detuning, as expected.  

To quantitatively fit the scans, we model the Rb vapor as an inhomogenously 
(Doppler) broadened two-level system \cite{Close1967}.  The general expressions 
for the absorptive and refractive coefficients involve complex error functions, 
but can be considerably simplified when the power-broadening of the homogenous 
linewidth is small compared to the Doppler width. In this limit the absorption 
coefficient and index of refraction are given by 

\begin{equation}
\alpha(x, I) =  \alpha_{0} \left( \frac{e^{-x^{2}}}{\sqrt{1+I/I_{s}}} \ - 
\frac{2a}{\sqrt{\pi}} \left( 1 - 2xF(x) \right) \right)
\label{absorptioneqn}
\end{equation}

\begin{equation} 
n(x, I) = 1 - \alpha_{0}\frac{ac}{\omega} \left( xe^{-x^{2}}\sqrt{1+I/I_{s}} \ 
- \frac{1}{a\sqrt{\pi}} \ F(x) \right)
\label{refractioneqn}
\end{equation}

\noindent Here $a$ is the homogeneous linewidth and $x$ is the detuning 
(positive for the red side), both normalized to the Doppler width.  $I$ is 
the beam intensity, $I_{s}$ is the atomic saturation intensity, $\omega$ is the 
laser angular frequency and $\alpha_{0}$ is the linear absorption coefficient in 
the absence of Doppler broadening. The function $F(x)$ is Dawson's integral, 
given by \cite{Abramowitz}

\begin{equation}
F(x) = e^{-x^{2}} \int_{0}^{x} e^{+t^2} \, dt.
\label{dawson}
\end{equation}

\noindent In the limit of low intensity, Eqs. \ref{absorptioneqn} and 
\ref{refractioneqn} become a Kerr (third-order) nonlinearity.

In a true Doppler-broadened two-level system, $\alpha_{0}$ and $I_{s}$ depend 
only on the homogeneous linewidth, resonance frequency, atomic density and 
dipole moment. Because of the hyperfine structure and the two isotopes $^{85}$Rb 
and $^{87}$Rb, our system is not truly two-level, so we allow $\alpha_{0}$ and 
$I_{s}$ to be parameters in our fit.  Our fitting procedure performs a 
numerical integration of the beam through the cell at each radial point, using 
Eqs. \ref{absorptioneqn} and \ref{refractioneqn}.  The output beam is then 
propagated to the far-field detector using a quasi-fast Hankel transform 
\cite{Siegman1977}. We use enough radial points so that, with absorption and 
refraction turned off, our model gives unity transmission to better than 0.05\%.  
To estimate the numerical error in our integration, we perform the fits twice, 
doubling the number of integration steps in the second instance.  The two 
fitting results differ by less than 2\%.  It is worth noting that although our 
cell is ``thin'' in the sense that it is much shorter than the beam Rayleigh 
range, it is not thin compared to the atomic absorption length.  This requires 
us to perform a full integration through the vapor, rather than simply imposing 
a single absorption and phase profile on the beam as it passes through the cell.

Our numerical fits give excellent agreement with the open-scan data, with 
parameters $I_{s}$ =  2.1 mW/cm$^{2}$ for both red and blue detuning, and 
$\alpha_{0}$ =  9.4 cm$^{-1}$, 9.9 cm$^{-1}$ for red and blue detuning, 
respectively (see Fig. \ref{openscan}).  To model the closed-aperture scans, we 
use these parameters to calculate the transmission vs. z curves that should 
result for a 20\% fluence aperture, finding good agreement within the systematic 
error (see Figs. \ref{closedscanblue}, \ref{closedscanred}).  The error is mostly 
from residual beam astigmatism and uncertainties about the aperture size.  We 
emphasize that we do not fit or adjust any parameters to produce the theory 
curves for the closed-aperture scans.  The agreement of the theory with the 
closed-aperture data is evidence of the Kramers-Kronig relations connecting the 
absorptive and refractive properties of this system, even though it is nonlinear 
\cite{textbooks}.  For low intensities, the fit parameters imply a Kerr 
coefficient ($n_{2}$) of -5.5 $\times$ 10$^{-6}$ cm$^{2}$/W for red detuning and 
+5.8 $\times$ 10$^{-6}$ cm$^{2}$/W for blue detuning. To check that we are in 
the limit in which Eqs. \ref{absorptioneqn} and \ref{refractioneqn} are valid, 
we note that the ratio of the power-broadened homogeneous linewidth (38 MHz 
times the saturation parameter $\sqrt{1 + I/I_{s}}$) to the Doppler width ($kv = 
2\pi \times$ 380 MHz, $v$ is the most probable atomic velocity) is 0.4 for the 
axial center of the beam at z = 0, falling off quadratically in z and 
exponentially in the beam radius.  Thus for calculations of the integrated 
transmission, Eqs. \ref{absorptioneqn} and \ref{refractioneqn} are good 
approximations.

To determine the effect of the unshielded geomagnetic field we repeated the 
experiment with vertical linear polarization, finding scans that differ by about 
10\%.  Our fits produced values of $\alpha_{0}$ that were the same as for 
horizontal polarization to within 5\%, but the $I_{s}$ fits were consistently a 
factor of 2 larger.  We attribute this to optical pumping effects, which can 
change the saturation intensity seen by the beam by changing the population of 
atoms in dark states.  With these altered parameters, the fit of the model to 
the data is excellent, indicating that a detailed inclusion of optical pumping 
in the model is not needed.

The success of this analytic, Doppler-broadened, two-level model in describing 
self-induced absorptive and refractive effects suggests that more complicated 
models are not needed to understand self-action in atomic vapor.  However, it is 
clear that this simple model fails to predict other important nonlinear effects 
such as cross-saturation and cross-phase modulation.  Their correct treatment 
requires more complicated models that include the effects of more atomic levels.

In conclusion, we have used both open and closed z-scans with powers of 
only tens of microwatts to measure the absorptive and refractive nonlinearities 
of Rb vapor within the Doppler profile of the D2 line.  Our open-aperture data 
are in excellent quantitative agreement with an analytic, Doppler-broadened 
two-level model.  Parameters from the open-aperture fits produce theory curves 
for the closed-aperture data that fit well to within the systematic error.  Our 
results imply that an analytic two-level model of the nonlinearity of Rb vapor 
is sufficient for describing self-action effects, as long as the powers involved 
do not broaden the homogeneous linewidth to larger than the Doppler width.

This work was supported in part by ONR and NSF Grant No. 0101501. JMH thanks the 
support from Instituto do Mil\^{e}nio de Informa\c{c}\~{a}o Qu\^{a}ntica, 
CAPES, CNPq, FAPEAL, PRONEX-NEON, ANP-CTPETRO.

\end{document}